\documentclass[fleqn,12pt]{article}
\usepackage{amsmath}
\usepackage{amssymb}
\usepackage[dvips]{graphicx}
\usepackage{bezier,amscd,amsmath}
\usepackage{amssymb}

\begin{document}

\title{Hyperbolic statics in space-time }
\author{D. G. Pavlov, Science Research Institute RSI HSGP (Fryazino), \\
S. S. Kokarev, RSEC ``Logos'\ (Yaroslavl)}
\date{}
\maketitle

\begin{abstract}
Based on the concept of material event as an elementary material source that
is concentrated on metric sphere of zero radius --- light-cone of Minkowski
space-time, we deduce the analog of Coulomb's law for hyperbolic space-time
field universally acting between the events of space-time. Collective field
that enables interaction of world lines of a pair of particles at rest
contains a standard 3-dimensional Coulomb's part and logarithmic addendum.
We've found that the Coulomb's part depends on a fine balance between causal
and geometric space-time characteristics (the two regularizations
concordance).
\end{abstract}

\bigskip

\section{Introduction}

Special Relativity (SR) formulated at the beginning of the XX century formed
a basis for the new understanding of space-time and physical processes
taking place in it. One of the key features of SR is its geometrical
interpretation: in the core of relativistic physics there lies the concept
of 4-dimensional Minkowski space-time $\mathcal{M}_{1,3}$ with
pseudo-Euclidean metric
\begin{equation*}
(\eta )=\text{diag}(1,-1,-1,-1).
\end{equation*}%
Geometric language of SR makes it possible to explicitly and consistently
formulate its essence and main foundations as well as to deduce its various
implications and make a consistent transition to General Relativity (GR).

Theoretical basis for this article is provided by the consequent space-time
interpretation of elementary objects in the  4-dimensional Minkowski world.
Newtonian laws were verified earlier \cite{kok1} within the concept of
4-dimensional statics of highly stretched strings. In this paper, we are
going to depart from the traditional interpretation of an elementary
particle and follow the logics of Minkowski 4-dimensional geometry. It means
that we take as an elementary physical object not a particle's world line,
but a true material point of $\mathcal{M}_{1,3}$ space --- metric sphere of
zero radius. In space-time it corresponds to a light-cone with material
characteristics concentrated on it. General considerations suggest that
these elementary object-sources correspond to a space-time field. Such field
is a 4-dimensional hyperbolic analog of Coulomb field. According to this
approach, the extended structures like world lines or world tubes stretched
along time-like direction should be obtained by the alignment (condensation)
of the elementary event points which can be described in frames of a certain
generalized theory of condensed media in 4-dimensional space-time\footnote{%
This theory would contain time-like forces and interactions that cannot be
found in standard relativistic physics.}. Naturally, there arises a question
about the relationship between their collective field and the observable
physical fields that are the subjects of standard physics. This question in
its simplest formulation is the main subject of this study.

\section{Hyperbolic solution with central symmetry}

We will use the hyperbolic analogue of Coulomb's law in Minkowski space as a
starting point for our considerations. This analogue can be defined as a
spherically symmetric (in the pseudo-Euclidean sphere sense) solution of the
wave equation
\begin{equation}
\Box U=0  \label{wave1}
\end{equation}%
in the empty space-time that surrounds the centre of the hyperbolic sphere.
At the same time, we keep in mind that the spherically symmetric solution of
Laplace equation in vacuum
\begin{equation}
\Delta \phi =0,  \label{laplas1}
\end{equation}%
is the unique one (up to a constant), while the immediate verification shows
that it contains all information about its point source. Indeed, the
solution of (\ref{laplas1}) in the form of Coulomb potential $\phi =q/r,$
actually, satisfies the equation
\begin{equation}
\Delta \phi =-4\pi q\delta (x)\delta (y)\delta (z)=-\frac{q\delta (r)}{r^{2}}%
.  \label{laplas2}
\end{equation}%
in all the space. The last equality takes into consideration the
transformation of delta-function when passing to curvilinear coordinates
\cite{vlad}.

As in case with Coulomb field, we don't put the question of the source
structure in equation (\ref{wave1}): the solution automatically contains the
singular characteristics of the source. In order to obtain this solution,
let us choose the hyperbolic spherically symmetric 4-dimensional coordinate
system with zero in the center :
\begin{equation}
\left\{
\begin{array}{lcr}
t & = & \varrho \cosh \chi ; \\
x & = & \varrho \sinh \chi \,\sin \theta \,\cos \varphi ; \\
y & = & \varrho \sinh \chi \,\sin \theta \,\sin \varphi ; \\
z & = & \varrho \sinh \chi \,\cos \theta .%
\end{array}%
\right.   \label{hypc}
\end{equation}%
Here $\varrho $ is the 4-radius, $\chi $ is the hyperbolic angle, $\theta $
and $\varphi $ are the pair of standard spherical angles. Formulae (\ref%
{hypc}) are valid for the domains where $t^{2}-x^{2}-y^{2}-z^{2}>0.$
Minkowski metric in this coordinate system can be obtained by using standard
rules for the transformation of an interval. It takes the following form:
\begin{equation}
ds^{2}=d\varrho ^{2}-\varrho ^{2}(d\chi ^{2}+\sinh ^{2}\chi (d\theta
^{2}+\sin ^{2}\theta d\varphi ^{2})).  \label{g1}
\end{equation}%
In differential geometry, the wave-operator can be invariantly determined by
the formula \cite{ward}:
\begin{equation}
\Box \equiv \frac{1}{\sqrt{-g}}\frac{\partial }{\partial \xi ^{\alpha }}%
\left( \sqrt{-g}g^{\alpha \beta }\frac{\partial }{\partial \xi ^{\beta }}%
\right) ,  \label{wave3}
\end{equation}%
where $g$ is the metric tensor determinant, $g^{\alpha \beta }$ is the
contravariant components of the metric, whose matrix is inverse to $%
(g_{\alpha \beta })$. From (\ref{g1}), it follows that $g=-\varrho ^{6}\sinh
^{4}\chi \sin ^{2}\theta ,$ while the inverse matrix takes the form:
\begin{equation}
(g^{\alpha \beta })=\text{diag}\left( 1,-\frac{1}{\varrho ^{2}},-\frac{1}{%
\varrho ^{2}\sinh ^{2}\chi },-\frac{1}{\varrho ^{2}\sinh ^{2}\chi \sin
^{2}\theta }\right) .  \label{wave4}
\end{equation}%
Substituting these expressions into the general formula (\ref{wave3}), we
obtain the expression for wave operator in the  4-dimensional spherical
coordinate system:
\begin{equation}
\Box =\frac{1}{\varrho ^{3}}\frac{\partial }{\partial \varrho }\left(
\varrho ^{3}\frac{\partial }{\partial \varrho }\right) -\frac{1}{\varrho ^{2}%
}\Delta _{\chi ,\theta ,\varphi },  \label{wave5}
\end{equation}%
where we've introduced the following designation for the angle part of wave
operator \footnote{%
When hyperbolic angles are small one have $\sinh \chi \approx \chi $ and
expression for $\Delta _{\chi ,\theta ,\varphi }$ transits into Laplace
operator in 3-dimensional spherical coordinate system with $r=\chi .$}:
\begin{equation}
\Delta _{\chi ,\theta ,\varphi }\equiv \frac{1}{\sinh ^{2}\chi }\left( \frac{%
\partial }{\partial \chi }\sinh ^{2}\chi \frac{\partial }{\partial \chi }+%
\frac{1}{\sin \theta }\frac{\partial }{\partial \theta }\sin \theta \frac{%
\partial }{\partial \theta }+\frac{1}{\sin ^{2}\theta }\frac{\partial ^{2}}{%
\partial \varphi ^{2}}\right) .
\end{equation}%
Substituting the general form of spherically symmetric solution $U=U(\varrho
)$ into operator \ref{wave5}), we obtain the following equation:
\begin{equation}
\Box U=\frac{1}{\varrho ^{3}}\frac{\partial }{\partial \varrho }\left(
\varrho ^{3}\frac{\partial U(\rho )}{\partial \varrho }\right) =0.
\label{wave7}
\end{equation}%
Its general, the solution takes the form:
\begin{equation}
U(\rho )=\frac{\mathcal{Q}}{\varrho ^{2}}+C,  \label{sol}
\end{equation}%
where $\mathcal{Q}$ and $C$ are integration constants.

Let us regard the obtained solution as an analogue of the fundamental
solution for the hyperbolic field whose sources are the \textit{material
events, }i.e. the\textit{\ }cone with the distributed characteristic $%
\mathcal{Q},$ which we will call the \textit{hyperbolic charge}. Immediate
verification shows that the obtained solution satisfies the 4-dimensional
analogue of the solution \footnote{%
It should be noted that technically it would be easier to write down the
right hand side in (\ref{sol1}) directly in spherically symmetric coordinate
system, because this form doesn't contain the infinite factor $\Omega _{H}$
, which is an analogue of the factor $4\pi $ in (\ref{laplas2}) which in its
turn determines the measure of set of all directions in $\mathcal{M}_{1,3}.$}
(\ref{laplas2}):
\begin{equation}
\Box U=-\frac{2\mathcal{Q}}{\varrho ^{3}}\delta (\varrho ).  \label{sol1}
\end{equation}%
Here we use the term "analogue of fundamental solution", because, unlike the
classical fundamental solution used in mathematical physics, the singularity
of (\ref{sol}) is concentrated on the light-cone and not in the point. We
will call the solution (\ref{sol}) \textit{hyperbolic fundamental solution}
of wave equation, in order to distinguish it from the well-known fundamental
solution (causal Green's function) of classical field theory:
\begin{equation}
G=\theta (t)\frac{\delta (t-r)}{4\pi r},  \label{green}
\end{equation}%
here $\theta (t)$ is Heaviside step function.

A simple analysis shows that the hyperbolic fundamental solution of the form
(\ref{sol}) satisfies the equation (\ref{sol1}) in generalized sense in all
causal domains.

\section{Static interaction of particles}

In order to reveal the intrinsic relation between the classical field theory
and the hyperbolic field, let us, first, consider the following simple
situation --- a pair of classical particle-sources at rest in a certain
inertial reference frame. In a 4-dimensional coordinate system adjusted to
this reference frame, the pair of these particles, described in $\mathcal{M}%
_{1,3}$ is the pair of world lines, parallel to the time-axis and separated
by spatial distance $r.$ These world lines are \textquotedblleft
weaved\textquotedblright\ \ out of material events. The superposition
principle for the hyperbolic field is valid, because of the linearity of the
wave equation. It means that the resulting field, $\phi ,$ of the world line
of particle 1 calculated for a certain point on the world line of particle 2
can be obtained by integration:
\begin{equation}
\phi (t_{2},r)=\int\limits_{-T/2}^{T/2}\frac{\lambda _{1}dt_{1}}{%
(t_{2}-t_{1})^{2}-r^{2}},  \label{sol2}
\end{equation}%
where $\lambda _{1}dt_{1}=d\mathcal{Q}_{1},$ $\lambda _{1}$ is the linear
density of hyperbolic charge 1, $T$ is the duration of particle history
(regularization parameter). Multipliing $\phi _{1}(t_{1},r)$ by element $%
\lambda _{2}dt_{2}$ of the hyperbolic charge of the second particle world
line, and  integrating along the same line, we obtain:
\begin{equation}
\phi _{12}(r)=\frac{1}{2}\int\limits_{-T/2}^{T/2}\int\limits_{-T/2}^{T/2}%
\frac{\lambda _{1}\lambda _{2}}{(t_{2}-t_{1})^{2}-r^{2}}dt_{1}dt_{2}.
\label{sol3}
\end{equation}%
This is the full energy of hyperbolic interaction between the classical
particles (factor $1/2$ appears because the double integration doubly
accounts for the same pair of elements on the world lines).

\medskip {\small Since the calculation of the integral involves two
regularizations, and each of them has certain physical meaning, the detailed
computations are shown below. By substituting variables $\xi _{1}=t_{1}/r,$ $%
\xi _{2}=t_{2}/r,$ we arrive to the factorization of dimensional and
dimensionless expressions in (\ref{sol3}):
\begin{equation}
\phi _{12}(r)=\frac{\lambda _{1}\lambda _{2}}{2}I(a),  \label{sol4}
\end{equation}%
where the dimensionless integral $I(a)$ depends only on dimensionless
parameter $a=T/2r,$ and it is expressed by the formula:
\begin{equation}
I(a)=\int\limits_{-a}^{a}\int\limits_{-a}^{a}\frac{d\xi _{1}\,d\xi _{2}}{%
(\xi _{2}-\xi _{1})^{2}-1}.  \label{sol5}
\end{equation}%
In the geometrical sense, it is the integral of an exact 2-form $(d\xi
_{1}\wedge d\xi _{2})/[(\xi _{2}-\xi _{1})^{2}-1]$ over the square domain $%
Q_{2a}.$ Let us now choose new coordinates: $u=\xi _{1}-\xi _{2};\quad v=\xi
_{1}+\xi _{2}.$ The area element  $d\xi _{1}\wedge d\xi _{2}=(du\wedge dv)/2,
$ and the integration domain on the plane of variables $(u,v)$ will appear
as a square $\bar{Q}_{2a}$ with vertices lying on the axis at the points
with coordinates $\pm 2a.$ Written in new variables, integral (\ref{sol5})
will take the form:
\begin{equation}
I(a)=\frac{1}{2}\int\limits_{\bar{Q}_{2a}}\frac{du\,dv}{u^{2}-1}.
\label{sol7}
\end{equation}%
In view of singularity of the integrand expression on straight lines $u=\pm
1,$ the regularization is required. The idea of the regularization described
below is to simultaneously cut out the contributions of $\epsilon $%
-neighborhoods of singular straight line segments ($\epsilon $-bands $%
B_{\epsilon 1}$ and $B_{\epsilon 2}$) to the integral, and then to perform
the limiting process $\epsilon \rightarrow 0.$ The boundaries of $\epsilon $%
-bands and the straight line $u=0$ (the 2-form on it is regular) fix the
following division of the integration domain:
\begin{equation}
\bar{Q}_{2a}=\Delta _{\epsilon 1}\cup B_{\epsilon 1}\cup T_{\epsilon 1}\cup
T_{\epsilon 2}\cup B_{\epsilon 2}\cup \Delta _{\epsilon 2},  \label{sol8}
\end{equation}%
where the triangular domains $\Delta _{\epsilon 1},\Delta _{\epsilon 2}$ are
specified by the following inequalities:
\begin{equation}
\Delta _{\epsilon 1}:\ -2a-u\leq v\leq 2a+u,\ \ -2a\leq u\leq -1-\epsilon ;
\label{sol9}
\end{equation}%
\begin{equation*}
\Delta _{\epsilon 2}:\ -2a+u\leq v\leq 2a-u,\ \ 1+\epsilon \leq u\leq 2a.
\end{equation*}%
The trapezoidal domains $T_{\epsilon 1},T_{\epsilon 2}$  are given by the
following inequalities:
\begin{equation}
T_{\epsilon 1}:\ -2a-u\leq v\leq 2a+u,\ \ -1+\epsilon \leq u\leq 0;\quad
\label{sol10}
\end{equation}%
\begin{equation*}
T_{\epsilon 2}:\ -2a+u\leq v\leq 2a-u,\ \ 0\leq u\leq 1-\epsilon .
\end{equation*}%
The integral (\ref{sol7}) in its regularized form now looks like this:
\begin{equation}
I(a,\epsilon )=\frac{1}{2}\int\limits_{\bar{Q}_{2a}\setminus (B_{\epsilon
1}\cup B_{\epsilon 2})}\frac{du\,dv}{u^{2}-1}.  \label{sol11}
\end{equation}%
In each of the regularization domains, the integral can be elementarily
calculated. After the computations and all the data collection, the result
is as follows:
\begin{equation}
I(a,\epsilon )=(2a-1)\ln (2a-1)-(2a+1)\ln (2a+1)+(2a+1)(\ln (2+\epsilon
)-\ln (2-\epsilon )).  \label{sum}
\end{equation}%
}

\medskip Passing to physical notations $2a=T/r$ in (\ref{sum}), we obtain:
\begin{equation*}
I(a,\epsilon )=
\end{equation*}%
\begin{equation}
\left( \frac{T}{r}-1\right) \ln \left( \frac{T}{r}-1\right) -\left( \frac{T}{%
r}+1\right) \ln \left( \frac{T}{r}+1\right) +\left( \frac{T}{r}+1\right)
(\ln (2+\epsilon )-\ln (2-\epsilon )).  \label{sum1}
\end{equation}%
If we pass to exact limits $T\rightarrow \infty ,$ $\epsilon \rightarrow 0$
in this expression, it will diverge independently on the order in which the
limiting processes are done. Let us consider a world in which the parameters
$T$ and $\epsilon $ differ from their ideal limit values. These parameters
have different physical meanings: the value of $T$ reflects "the duration of
history"\thinspace\ of the particle-sources, while the value of $\epsilon $
represents causality. If $\epsilon =0,$ the interaction due to the
hyperbolic field propagates strictly at the speed of light along the cones.
Small deviations of $\epsilon $ from zero correspond to the picture in which
the cones are slightly \textquotedblleft blurred\textquotedblright . Thus,
parameter $\epsilon $ acquires the meaning of an additional "fundamental
variable" \thinspace\ which could be formally described as
\begin{equation}
\epsilon =\delta c/c,  \label{sum2a}
\end{equation}%
where $\delta c$ is the absolute undeterminacy of the speed of light $c$
("fundamental constant"\thinspace ). In order to describe such a world which
has more general properties than Minkowski space-time has in SR, it would be
natural to start by considering not the limit of expression(\ref{sum1}),
when $T\rightarrow \infty ,$ $\epsilon \rightarrow 0,$ but rather its
asymptotic form under these conditions. Restricting ourselves by a couple of
first terms of the corresponding expansions (and cutting inessential
additive constants), we obtain:
\begin{equation}
I(a,\epsilon )\overset{\text{as}}{=}\frac{\epsilon T}{r}+2\ln r+O(\epsilon
^{3})+O((T/r)^{-5}).  \label{sum2}
\end{equation}%
The expression (\ref{sum2}) can be divided into logarithmic part and Coulomb
part. The latter is retained due to the finite value of the product $%
\epsilon T,$ which is responsible for a specific balance between the history
duration and causality.

Taking into account (\ref{sol4}), the asymptotic approximation of the final
expression for the interaction energy of the pair of particle-sources at
rest is as follows:
\begin{equation}
\phi _{12}(r)\overset{\text{as}}{=}\frac{\alpha _{1}\alpha _{2}}{r}+\lambda
_{1}\lambda _{2}\ln r,  \label{result}
\end{equation}%
where $\alpha _{i}=\lambda _{i}\sqrt{\epsilon T/2}$ are the Coulomb charges,
$\lambda _{i}$ are the logarithmic charges that coincide with linear density
of the original hyperbolic charge. The graph of function (\ref{result}) is
shown on fig.\ref{fig1}.

{\centering\leftskip0em\rightskip5em{\small \refstepcounter{figure}\label%
{fig1} \includegraphics[width=0.5\textwidth]{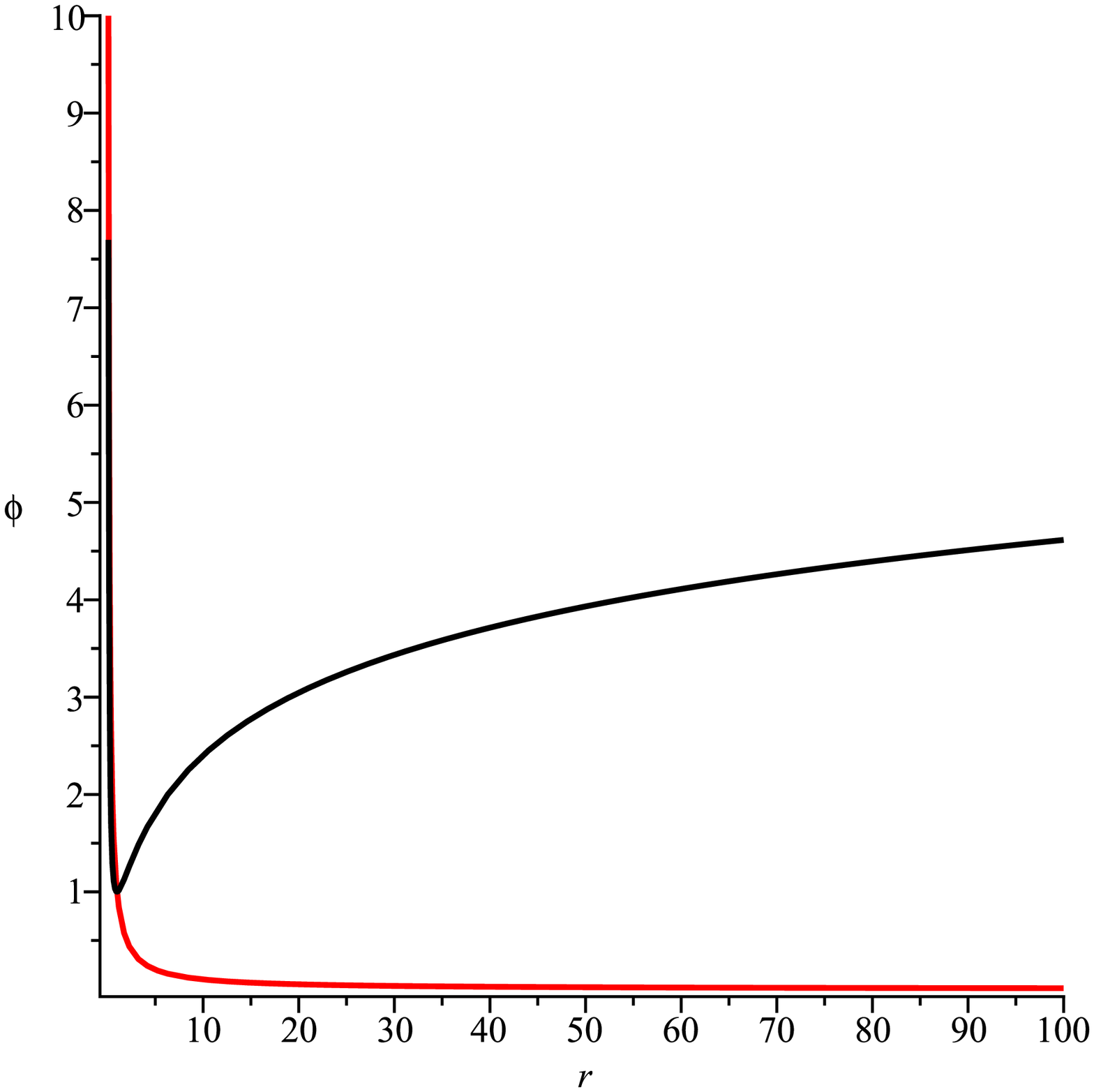} \medskip \nopagebreak}}

{\small 8A.~\thefigure. Energy of hyperbolic particle interaction. The black
curve is the potential (\ref{result}) with $\lambda _{1}=\lambda _{2}=1,$ $%
\alpha _{1}=\alpha _{2}=1,$ the red curve is its Coulomb part.}

\medskip

The Coulomb part of the potential (\ref{result}) dominates for small
distances, the logarithmic one dominates for large distances. According to
the asymptotic theory described here, the "small distances" are defined by
the natural condition $r\ll \epsilon T.$ To make a rough concordance with
observations, let us set $\epsilon \lesssim 10^{-10}$ (present accuracy of
light speed measurement), $T\gtrsim 10^{24}$s (today's notion for the time
scale of Universe existence). Then the Coulomb domain can be determined by
the inequality: $r\ll 10^{14}$m, which excedingly covers the solar system
scale. On the other hand, the logarithmic part inevitably dominates on
cosmological scales. As can be easily demonstrated, the logarithmic
potential naturally ensures the flat character of rotation curves, related
to common massive center, making it unnecessary to refer to the dark matter
concept. Indeed, Newton's second law for the attractive force in rotary
motion is  $\sim 1/r,$ and we obtain:
\begin{equation}
\frac{v^{2}}{r}\sim \frac{A}{r}\Rightarrow v\sim \text{const}.
\label{kepler}
\end{equation}

We would like to stress that our approach eliminates the need for the dark matter concept in principle,
as it is also the case with theories like MOND (Modified Newton Dynamics) \cite{mond}
or AGD (Anisotropic Geometrodynamics) \cite{s1,coll}.
Contrary to MOND, we do not modify the form of Newton second law, and unlike AGD,
we do not rely mainly on geometry. Our approach deals with the principally new definition
of an elementary physical object and with the modified fundamental law of interaction between such objects.

\bigskip \bigskip

{\large\bf Acknowledgements}

\bigskip

We are grateful to recently deceased academician of RAS, prof. V. G. Kadyshevskiy for the long-term interest in the topic of the paper and to prof. S.V. Siparov for helpful discussions.

\end{document}